\documentclass[%
reprint,
superscriptaddress,
amsmath,amssymb,aps,
]{revtex4-2}

\usepackage{units}

\usepackage{graphicx}
  \graphicspath{{pictures/}}
\usepackage{xcolor}

\begin{document}

\title{Metahouse: noise-insulating chamber based on periodic structures}

\author{Mariia Krasikova}
\email{mariia.krasikova@metalab.ifmo.ru}
\affiliation{School of Physics and Engineering, ITMO University, St. Petersburg 197101, Russia}

\author{Sergey Krasikov}
\affiliation{School of Physics and Engineering, ITMO University, St. Petersburg 197101, Russia}

\author{Anton Melnikov}
\email{anton.melnikov@tum.de}
\affiliation{Chair of Vibroacoustics of Vehicles and Machines, Technical University of Munich, Garching b. München 85748, Germany}

\author{Yuri Baloshin}
\affiliation{School of Physics and Engineering, ITMO University, St. Petersburg 197101, Russia}

\author{Steffen Marburg}
\affiliation{Chair of Vibroacoustics of Vehicles and Machines, Technical University of Munich, Garching b. München 85748, Germany}

\author{David Powell}
\affiliation{School of Engineering and Information Technology, University of New South Wales, Northcott Drive, Canberra, Australian Capital Territory 2600, Australia}

\author{Andrey Bogdanov}
\affiliation{School of Physics and Engineering, ITMO University, St. Petersburg 197101, Russia}

\begin{abstract}
Noise pollution remains a challenging problem requiring the development of novel systems for noise insulation. Extensive work in the field of acoustic metamaterials has led to the occurrence of various ventilated structures which, however, are usually demonstrated for rather narrow regions of the audible spectrum. In this work, we further extend the idea of metamaterial-based systems developing a concept of a \textit{metahouse} chamber representing a {\it ventilated} structure for {\it broadband noise insulation}. Broad stop bands originate from strong coupling between pairs of Helmholtz resonators constituting the structure. We demonstrate numerically and experimentally the averaged transmission $-18$~dB within the spectral range from $1500$ to $16500$~Hz. The sparseness of the structure together with the possibility to use optically transparent materials suggest that the chamber may be also characterized by {\it partial optical transparency} depending on the mutual position of structural elements. The obtained results are promising for development of novel noise-insulating structures advancing urban science.
\end{abstract}

\maketitle

\section{Introduction}
Acoustic metamaterials and phononic crystals have been subjects of keen scientific interest for a couple of  decades as they provide novel routes towards efficient control of acoustic field~\cite{cummer2016controlling,zangeneh-nejad2019active,liu2020review,liao2021acoustic}. In addition to the realization of negative refractive index~\cite{kaina2015negative}, superlenses~\cite{derosny2002overcoming,li2009experimental}, holograms~\cite{zhang2020acoustic}, and acoustic cloaks~\cite{chen2007acoustic}, recent advances include the development of non-reciprocal systems~\cite{guo2019nonreciprocal}, topological insulators~\cite{peng2016experimental,xue2020observation}, nonlinear~\cite{fang2017ultralow}, tunable~\cite{chen2018review}, coding~\cite{xie2017coding}, and programmable metasurfaces~\cite{tian2019programmable}. Acoustic metasurfaces were also explored as potential platforms for analog computing~\cite{zuo2018acoustic} and, vice versa, advances in computer science and artificial intelligence boost design procedures to achieve desired properties of metamaterials and metasurfaces~\cite{donda2021ultrathin,xu2021machinelearningassisted,gao2021inverse,ding2021deep}.
Metamaterials can be also used as a platform to explore analogies of the quantum concepts, such as Hall effect~\cite{zhou2020voltagecontrolled,fan2021acoustic}, spin properties~\cite{shi2019observation, long2020realization,yang2021real,wang2021spinorbit}, skyrmions~\cite{ge2021observation}, and twistronics~\cite{gardezi2021simulating}.

One of the developing branches in the field of acoustic metamaterials is dedicated to realization of novel noise-insulating systems. Increasing noise pollution in urban areas is one of the endangering trends affecting global health and ecological environment~\cite{stansfeld2003noise,singh2004noise,goines2007noise,kunc2019effects,munzel2021transportation}. Solution of this problem requires the development of new methods and materials allowing broadband passive noise insulation. Conventionally used systems are typically represented by bulky structures imposing harsh engineering restrictions on buildings and constructions~\cite{hopkins2007sound}. Frequency range of noise mitigation has to be compromised with mass and volume of the used materials. Moreover, some crucial properties like ventilation or optical transparency are usually incompatible with such systems.

\begin{figure}[htbp!]
    \centering
    \includegraphics[width=\linewidth]{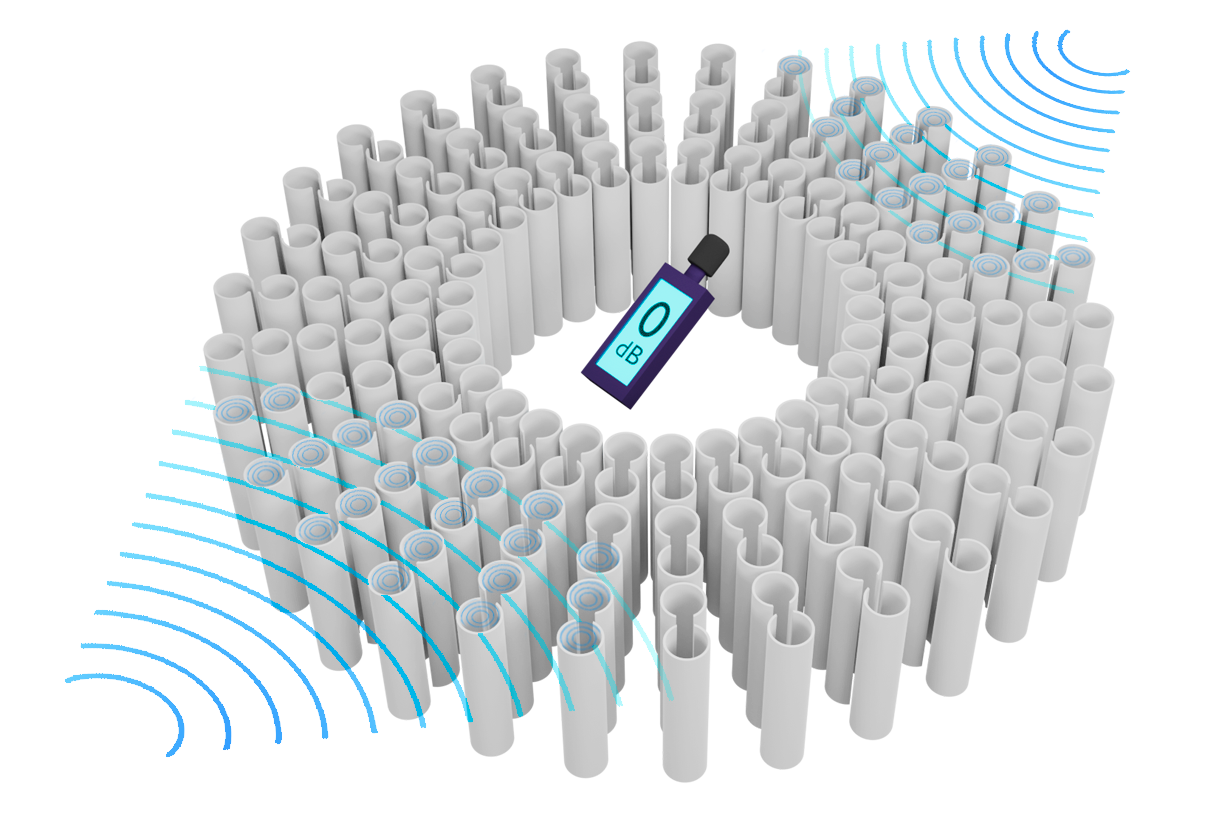}
    \caption{{\bfseries \textit{Metahouse} concept}. The Helmholtz resonators are shown with gray only for visual clearance and can be made of transparent materials. The inner space of the developed chamber is protected from acoustic noises within a broad audible spectral range. At the same time, the structure itself is ventilated and optically transparent. Two former parameters depend solely on the distance between the structural elements and their size, which also determine the spectral range of blocked noises. }
    \label{fig:concept}
\end{figure}

Instead of the conventional mass-density law, reflection and attenuation of sound in metamaterials relies mostly on periodicity and shape of the structural elements rather than their material properties. One of the important options enabled by metamaterials is the possibility to implement structures allowing air flow~\cite{xu2020topologyoptimized,sun2020broadband,kumar2020ventilated,dong2021ultrabroadband}. A variety of designs includes perforated membranes~\cite{ma2013lowfrequency, langfeldt2017perforated}, space coiling structures~\cite{cheng2015ultrasparse, chen2015openstructure, zhang2017omnidirectional, ghaffarivardavagh2019ultraopen, yu2019sound}, and metacages~\cite{shen2018acoustic, melnikov2020acoustic} have been presented.
Nevertheless, despite the plethora of achievable physical effects, acoustic metamaterials rarely find real-life applications. These structures usually have complicated designs and a narrow operating range.

In this paper, we propose a noise-insulating ventilated {\it \textit{metahouse} chamber} allowing access of light into the inner area. The chamber is designed to be simple for manufacturing and assembling. At the same time, the requirements for materials are easily met and the structure can be made of ecologically friendly and optically transparent materials.
The chamber is based on coupled Helmholtz resonators with C-shaped cross-section and exploits  band-gap formation as the main mechanism for sound insulation. The artistic illustration of the concept is presented in Fig.~\ref{fig:concept}. It is proposed that such structures may be incorporated into urban parks to create noise-free relaxation zones for visitors.

We start with designing a 2D metamaterial-based periodic structure and tuning its parameters for maximization of the integral width of band-gaps in the audible spectral range. Then, we perform an experimental demonstration of stop-bands corresponding to band-gaps of the infinite system. Finally, we provide an experimental demonstration of a finite-size prototype of the \textit{metahouse} chamber. All experimental results are verified by numerical simulations.

\section{Results}
\label{sec:results}

\subsection{Design of a unit cell}
\begin{figure}[htbp!]
    \centering
    \includegraphics[width=\linewidth]{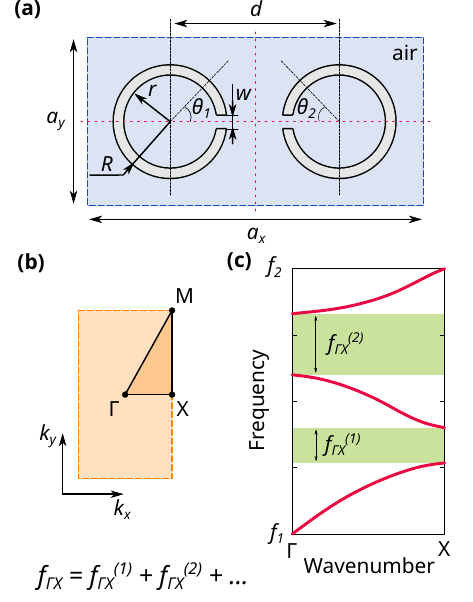}
    \caption{{\bfseries Design of the unit cell.}
    (a) Illustration of the considered unit cell consisting of two locally coupled Helmholtz resonators. Width of the unit cell is $a_x$ and the height is $a_y$. The resonators are characterized by the inner radius $r$ and the outer radius $R$. Slit width $w$ and distance between the resonators $d$ are tuning parameters. The resonators are rotated around their out-of-plane axes by the angles $\theta_1$ and $\theta_2$. Dashed red lines indicate reflection symmetry axes of the cell. 
    (b) The corresponding reciprocal unit cell. 
    (c) Arbitrary band diagram representing labeling of individual $\Gamma X$ band-gaps used for calculation of the integral width of $\Gamma X$ band-gaps $f_{\Gamma X}$.}
    \label{fig:cell_design}
\end{figure}

The chosen unit cell consists of two coupled Helmholtz resonators with C-shaped cross-section, as shown in Fig.~\ref{fig:cell_design}(a). Such resonators can be 3D printed or easily manufactured via simple mechanical processing of commercially available polymer pipes.
Similar geometries were already considered for noise mitigation purposes~\cite{melnikov2019acoustic,melnikov2020acoustic,elford2011matryoshka} and we have just utilized the fact that local coupling between the resonators may increase the integral width of band gaps~\cite{hu2017acoustic}. In this paper we consider only $\Gamma X$ band-gaps since the implemented experimental realization of the semi-infinite structure implies that the incident plane wave has only $k_x$ component. 
Performing the tuning of geometrical parameters, we have optimised the system to further increase the integral width of band-gaps $f_{\Gamma X}$ defined as the sum of the widths of individual band-gaps [see Figs.~\ref{fig:cell_design}(b) and~\ref{fig:cell_design}(c)]:
\begin{equation}
  f_{\Gamma X} = \sum f_{\Gamma X}^{(i)}
  \label{eq:total_bg_width}
\end{equation}
where $i$ stays for the number of a band-gap and index $\Gamma X$ indicates that the considered band-gaps lie within $\Gamma X$ interval of the Brillouin zone [see Fig.~\ref{fig:cell_design}(b) and~\ref{fig:cell_design}(c)]. Here it should be noted that summation is done over the band-gaps within the considered frequency ranges.
For simplicity, we have considered a 2D geometry assuming that the resonators are infinite in the direction perpendicular to the surface of the cross-section.

Since the system is scalable, for the tuning procedure we have normalized all geometric parameters to the outer radius of the resonators $R$. The size of the unit cell was fixed at values of $a_x/R = 4.5$ and $a_y/R = 2.25$.
The inner radius of the resonators $r$ is fixed as $r/R = 0.905$. The distance between the centers of resonators $d$ as well as the width of slits $w$ were varied as parameters directly affecting the local coupling. The frequencies are also normalized by $R/c_{\mathrm{air}}$, where $c_{\mathrm{air}} = \unit[343]{m/s}$ is the speed of sound in air.

\begin{figure*}[htbp!]
    \centering
    \includegraphics[width=\linewidth]{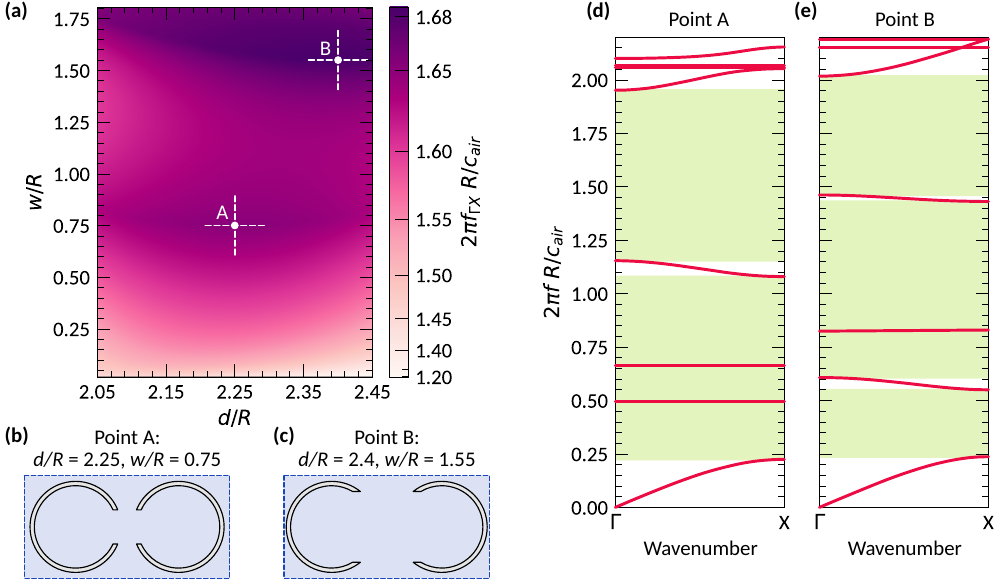}
    \caption{{\bfseries Band structure and optimization.} (a) Integral width of band-gaps as a function of slit width $w$ and distance between the resonators $d$. Points A and B indicate two local maxima of the function $f_{\Gamma X}$, occurring near the points with $w/R = 0.75$, $d/R = 2.25$ and $w/R = 1.55$, $d/R = 2.4$, respectively. The corresponding unit cells are shown in panels (b) and (c). Note that the map is rescaled using a power-law relationship 
    to provide better visualization.
    Band diagrams for the structures with such configurations are demonstrated in panels (d) and (e). Shaded green areas indicate the $\Gamma X$ band-gaps.}
    \label{fig:tuning}
\end{figure*}

The colormap shown in Fig.~\ref{fig:tuning}(a) demonstrates the integral width of $\Gamma X$ band-gaps $f_{\Gamma X}$ as a function of $w$ and $d$. Clearly, the presence of finite-size slits results in the increase of $f_{\Gamma X}$. At the same time, the dependence is nonlinear and there are two local maxima corresponding to the points with $d/R = 2.25$, $w/R = 0.75$ [point A, see Fig.~\ref{fig:tuning}(b)] and $d/R = 2.4$, $w/R = 1.55$ [point B, see Fig.~\ref{fig:tuning}(c)]. The corresponding band structures demonstrating large $\Gamma X$ band-gaps are shown in Fig.~\ref{fig:tuning}(d) and~\ref{fig:tuning}(e), respectively. 
It is also important to note that within the considered frequency range the structure may possess the features of a metamaterial as well as a phononic crystal. So there may be several mechanisms of band-gap formation involved.

\begin{figure}[htbp!]
    \centering
    \includegraphics[width=\linewidth]{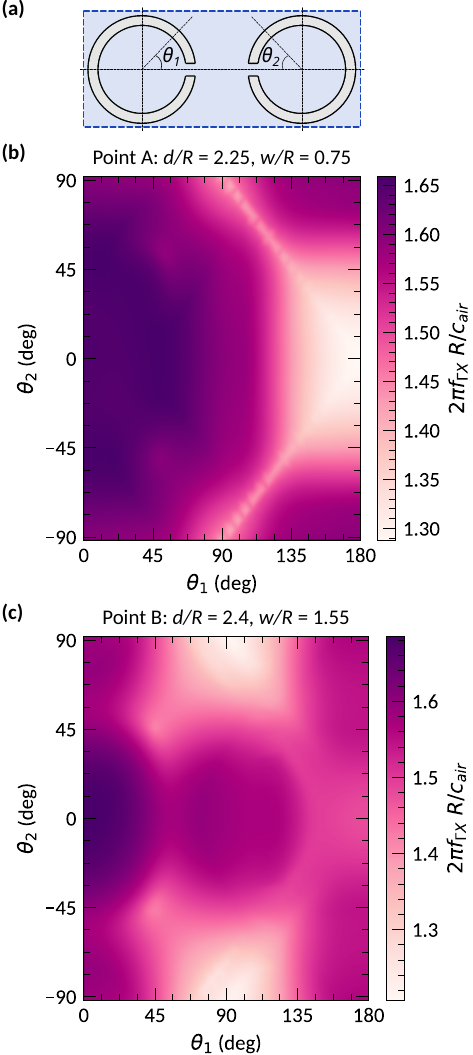}
    \caption{{\bfseries Rotation of the resonators.} The panel (a) illustrate the definition of the rotational angles $\theta_1$ and $\theta_2$. The strongly coupled geometry is characterized by $\theta_1 = \theta_2 = 0^{\circ}$ while the case of the CC-like geometry corresponds to $\theta_1 = 0^{\circ}$, $\theta_2 = 180^{\circ}$.
    The colormaps (b) and (c) show the integral width of band-gaps as a function of the rotational angles such that the resonators are characterized by (b) $w/R = 0.75$, $d/R = 2.25$ and (c) $w/R = 1.55$, $d/R = 2.4$. These parameters correspond to the points labeled in Fig.~\ref{fig:tuning} as A and B, respectively.}
    \label{fig:rotation}
\end{figure}

Following the adjustment of slit widths and distances between the resonators, we analyze the effect of resonators' rotation on the integral width of band-gaps. Figure~\ref{fig:rotation} shows the integral width of band-gaps calculated for the range of rotational angles $\theta_1$ and $\theta_2$ [see Fig.~\ref{fig:rotation}(a)]. Rotation of the resonators results in a pronounced decrease of $f_{\Gamma X}$ in the vicinity of the point $\theta_1 = 0^{\circ}$, $\theta_2 = 180^{\circ}$, corresponding to the case of CC-like geometry. Such a reduction of the integral width of band-gaps is associated with the decrease of local coupling strength~\cite{hu2017acoustic} which in the case of CC-like geometry is weak. In the Supplementary Information we demonstrate that the proposed geometry is indeed characterized by much stronger local coupling allowing increase of integral width of band gaps by nearly \unit[30]{\%} comparing to the CC-like geometry.  Considering the point A, for CC-like geometry $f_{\Gamma X}$ reduces by more than \unit[20]{\%} in comparison to the optimal case with $\theta_1 = \theta_2 = 0^{\circ}$ [see Fig.~\ref{fig:rotation}(b)]. Similarly, for point B the corresponding reduction is about \unit[15]{\%} [see Fig.~\ref{fig:rotation}(b)], however, there are configurations which are less beneficial.
Figure~\ref{fig:rotation} also demonstrates the robustness of the considered system against rotations of the resonators. Indeed, the calculations indicate that rotation within approximately $15^{\circ}$ almost does not perturb values of $f_{\Gamma X}$. This fact is important for practical realization of the proposed structure since it allows to reduce requirements for mounting tolerances.

\subsection{Experimental verification of stop-bands}
\begin{figure*}[htbp!]
  \centering
  \includegraphics[width=\linewidth]{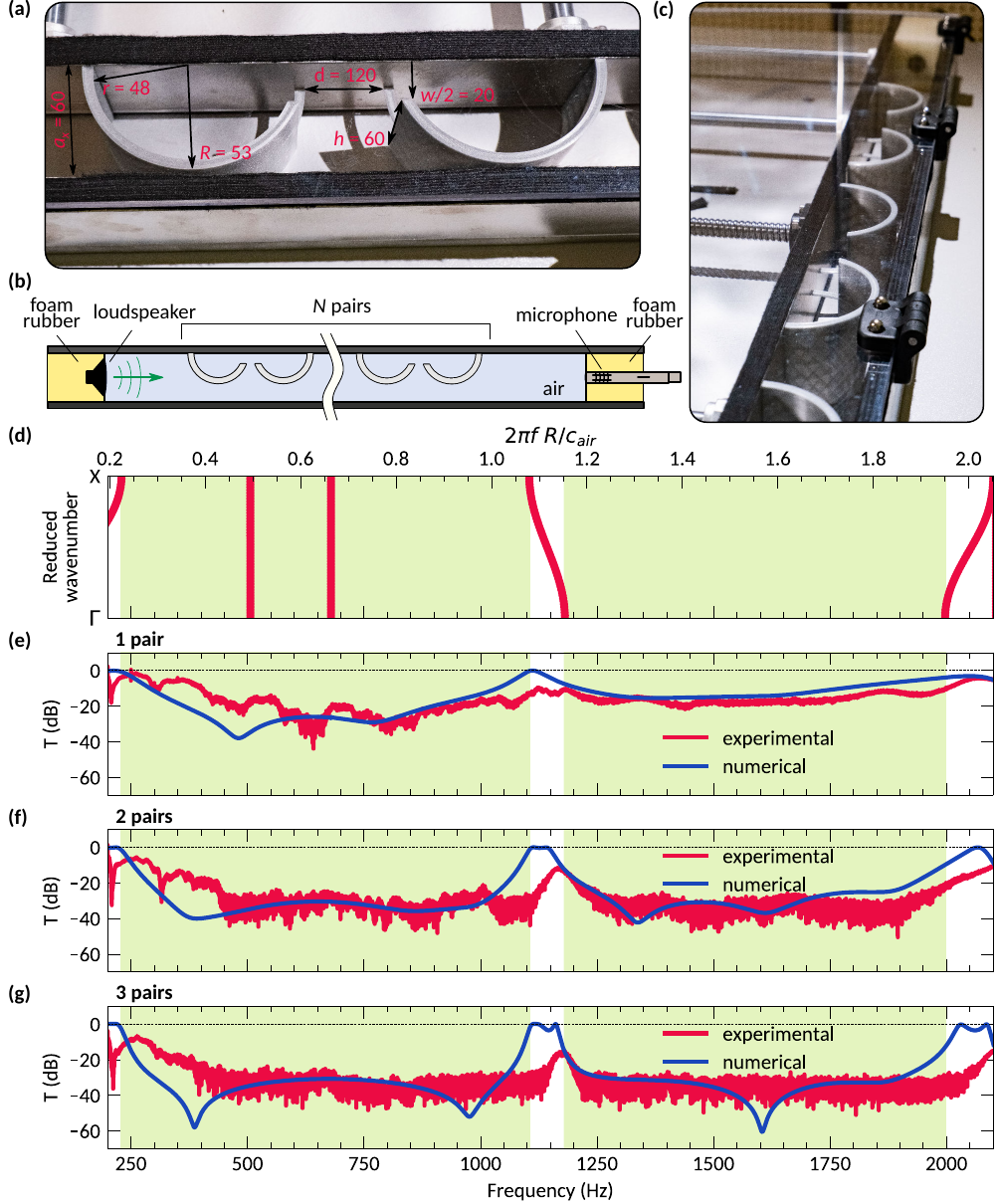}
  \caption{{\bfseries Transmission of semi-infinite structure.} (a) Photograph of the samples characterized by $R = 53$~mm, $r = 48$~mm, $w = 40$~mm, and height $h = 60$~mm. (b) Schematic picture and (c) photo of the experimental setup.
  (d) Band diagram for the system with the unit cell with $a_x = 240$~mm, $a_y = 120$~mm and $d = 120$~mm. For convenience, two frequency axes are shown, the normalized one and the one in conventional units. Measured and simulated (for a 2D system) transmission spectra for (e) one, (f) two, and (g) three pairs of coupled Helmholtz resonators. Experimentally obtained values are shown by red lines, while the numerical results are demonstrated by the blue lines. Green shaded areas indicate $\Gamma X$ band-gaps of the corresponding infinite 2D structure.}
  \label{fig:semi_infinite}
\end{figure*}

To experimentally verify the presence of wide band-gaps, we consider a configuration of the structure corresponding to the point A in Fig.~\ref{fig:tuning}. We imitate the semi-infinite structure, such that it is finite along one axis and infinite along the two other ones (see Methods section for details).
The samples fabricated via 3D printing are shown in Fig.~\ref{fig:semi_infinite}(a) while the schematic view and photograph of the setup are demonstrated in Figs.~\ref{fig:semi_infinite}(b) and~\ref{fig:semi_infinite}(c), respectively. Note that of instead the complete resonators we consider their halves utilizing the symmetry of the field distribution within the unit cell, as discussed in Methods section.
The measurements were provided for the unit cell characterized by $a_x = \unit[240]{mm}$ and $a_y = \unit[120]{mm}$ such that the outer radius of the resonators is $R = \unit[53]{mm}$ and the inner radius is $r = \unit[48]{mm}$. Slit width is then $w = \unit[40]{mm}$ and the distance between the resonators is $d = \unit[120]{mm}$. The considered frequency range is \unit[200 -- 2060]{Hz}.

The obtained transmission spectra are demonstrated in Figs.~\ref{fig:semi_infinite}(e) --~\ref{fig:semi_infinite}(g), where the blue lines correspond to numerical simulations (for a 2D system) and the red lines represent the experimental measurements. The green shaded areas indicate the $\Gamma X$ band-gaps of the equivalent infinite 2D system shown in Fig.~\ref{fig:semi_infinite}(a). For convenience, two frequency axes are shown, the normalized (top) and the un-normalized (bottom).
The measurements are provided for $1$, $2$, and $3$ pairs of resonators in order to explicitly demonstrate the evolution of stop-bands with respect to a change in the structure thickness.
In accordance with a simple intuition, increasing the number of pairs results in broadening of the stop-bands.
It should be noted that, unlike the noise barrier consisting of a single Helmholtz resonator~\cite{melnikov2020acoustic}, a wide stop-band can be observed even for the one-pair thickness.
Practically, for the case of $3$ pairs of coupled resonators, the stop-bands cover more than \unit[80]{\%} of the considered spectral range. But even for one pair the stop-bands are quite large and pronounced, mostly going below $-15$~dB.

\subsection{Noise-insulating chamber}
\begin{figure*}[htbp!]
    \centering
    \includegraphics[width=\linewidth]{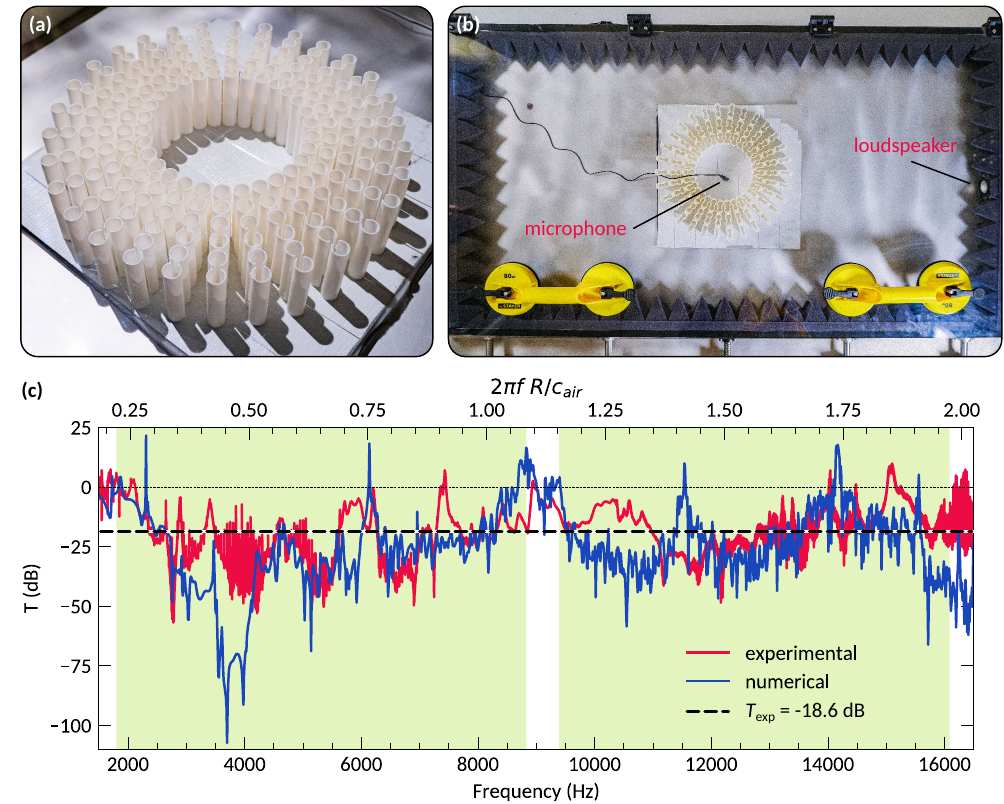}
    \caption{{\bfseries Transmission of the \textit{metahouse} chamber.} (a) The manufactured prototype of the \textit{metahouse} chamber. There are $30$ arrays forming a toroid with the outer radius $300$~mm and the inner one $120$~mm.  Each array consists of $3$ pairs of resonators with $R = 6.7$~mm, $r = 5.5$~mm, and $w = 5$~mm. The distance between centers of the resonators along the axis of an array is $d = 15$~mm. (b) Photograph of the experimental setup. The microphone is located inside the chamber and the loudspeaker is placed at one of the walls. (c) Measured transmission spectrum (red line) and the corresponding numerical values obtained for the equivalent 2D chamber (blue line). Shaded green areas correspond to the equivalent infinite 2D periodic structure [see Fig.~\ref{fig:tuning}].}
    \label{fig:metacage}
\end{figure*}

The finite-size prototype of the \textit{metahouse} chamber is shown in Fig.~\ref{fig:metacage}(a). It is an axially symmetric structure consisting of $30$ radially oriented arrays with $3$ pairs of resonators each. The outer radius of the chamber is $300$~mm and the inner one is $120$~mm. Height of the structure is $60$~mm. The resonators are characterized by $R = 6.7$~mm, $r = 6$~mm, and $w = 5$~mm. Along the arrays, the distance between centers of the resonators (in $xy$ plane) is $d = 15$~mm. 
The transmission spectrum is shown in Fig.~\ref{fig:metacage}(c) with experimental values indicated by red line and the numerical simulations (for a 2D system) by the blue one. Shaded green areas indicate $\Gamma X$ band-gaps of the corresponding infinite 2D system considered in the previous Section.

The measurements were performed in a box with the walls covered by foam rubber pyramids [see Fig.~\ref{fig:metacage}(b)]. However, it has to be noted that the height of the box is $60$~mm while the wavelength of generated waves is smaller than this value for almost whole range of spectra (see Methods section). Since all of the structures are uniform in the vertical direction, it is unlikely that the higher order modes would be excited. Still, assuming that the system is not ideal it may be not a perfect equivalent of the 2D system which explains some differences between the numerical and experimental results.

\section{Discussion}
The developed \textit{metahouse} chamber demonstrates good insulating properties with the average transmission of $-18$~dB in the broad spectral range from $1500$ to $16500$~Hz [Fig.~\ref{fig:metacage}(c)]. The average value was calculated as the arithmetic mean within the spectral range. Ventilation properties as well as partial optical transparency are defined by geometric properties of the structure. Analysis of the air flow is presented in the Supplementary Information.
While the concept was demonstrated for the case when the resonators are located quite close to each other, the structure may be optimized to increase the sparseness. For example, numerical calculations done for the finite-size metacage with the number of arrays reduced by a half also demonstrate reasonable noise-insulation properties with the average transmission $-7$~dB (see Supplementary Information). Tuning of parameters and optimization of the structure may enhance the results allowing proper air flow and optical transparency implying compromise between sparseness and sound transmission. In addition, the structure can be made of optically transparent materials, such as the commercially available PLA filament for 3D printers~\cite{matos2019evaluation}. The possibility to achieve simultaneous air ventilation with optical transparency is a unique feature of the developed structure resulting from the proposed design. In Table~\ref{tab:impedances} we also provide values of acoustic impedance for several materials which may be used for construction of the \textit{metahouse} chamber. It is important to highlight that the range of possible materials is huge and hence the conventional widespread materials can be used.

\begin{table}[htbp!]
    \centering
    \caption{Acoustic impedance $Z$ of some materials which can be used for construction of the \textit{metahouse} chamber.}
    \label{tab:impedances}
    \begin{tabular}{|c|c|c|}
        \hline
        Material & $Z$ (MRayl) & Reference \\
        \hline
        \hline
        Polylactic acid (PLA) & 2.56 -- 2.77 & \cite{parker2010longitudinal} \\
        Polypropylene & 2.3 -- 2.4 & \cite{selfridge1985approximate} \\
        Polymethyl methacrylat & 3.14 -- 3.69 & \cite{carlson2003frequency} \\
        \hline
    \end{tabular}
\end{table}

\section{Methods}
\label{sec:methods}

\subsection{Numerical simulations}
All numerical simulations are performed for 2D systems using finite element method (FEM) via COMSOL Multiphysics, Pressure Acoustics module. Throughout the calculations, the boundaries of resonators are assumed to be sound-hard walls. Such an approximation is justified by high impedance contrast between the material of resonators and air. For example, the acoustic impedance of polypropylene is about $\unit[2.4]{MRayl}$~\cite{selfridge1985approximate} which is several orders higher than for air. Similar contrasts of impedances is also observed for other materials which can be used for construction of the chamber, see Table~\ref{tab:impedances}.
It is also should be mention that any thermo-viscous boundary effects are neglected, since there are no narrow channels in the considered structures.

The semi-infinite structure is modelled as an array made of $N$ halved resonators placed into a 2D waveguide with sound hard walls. The distance between the walls is equal to $a_y/2$. The ends of the waveguide are modelled as perfectly matched layers (PML). The incident plane wave with amplitude $p_0$ propagates along the direction of periodicity ($x$-axis, parallel to the walls of the waveguide). Pressure values are calculated at the sample point located at the distance $x_t = 15$~mm from the rightmost resonator. Parameters of the resonators correspond to those of the fabricated samples.

In the model for the finite-size case the chamber is placed into the 2D box in which all walls are PML such that the distance between the walls is $1210$~mm and $510$~mm along $x$ and $y$ axis, respectively. The structure is located at the center of the waveguide and the sample point is placed at the center of the structure. Incident waves with amplitude $p_0$ are generated by a point source located near one of the walls. Geometric size of the 2D chamber correspond to the parameters of the manufactured structure.

\subsection{Experimental setup}
The experimental measurements are provided inside a waveguide with rectangular cross-section. The walls and the bottom of the waveguide are made of aluminum with thickness of $15$~mm. The lid is made of $6$~mm thick transparent plexiglass. The height of the inner cross-section of the waveguide is fixed and equal to $60$~mm. However, the width is adjustable and can be varied from $0$ to $650$~mm via displacement of one of the walls. 
The loudspeaker (Visaton BF 45/4) is located at one end of the waveguide and the microphone (BOYA BY-PVM1000) is placed at the other one. While the full length of the waveguide is \unit[135]{cm}, the distance between the microphone and the loudspeaker is \unit[107]{cm}. The rest of the waveguide's length is occupied by two pieces of foam rubber, each \unit[14]{cm} long. The loudspeaker and the microphone are embedded into these inserts. 

The microphone is directly connected to a USB audio interface Roland Rubix22. To connect the loudspeaker to the same sound card we use an amplifier based on the Yamaha YDA138-E microchip. The sound card itself is connected to a personal computer via USB interface. Control of signal generation and recording is realized via specially developed software based on the Python programming language (in particular, generation and recording of signals exploit the sound device module~\cite{2021play}).

To imitate semi-infinite structure in the experiment we consider an array consisting of $N$ pairs of coupled resonators. Proper boundary conditions --- namely, sound hard walls at geometrical symmetry axes --- allow to imitate periodicity along the two other axes ($y$ and $z$ axes). The numerical 2D model corresponds to a structure infinite in the $z$ direction, while in the experiment it is mimicked by keeping the  waveguide width and height smaller than $\lambda/2$ for the considered spectral range.
The operational range of the setup for the case of the semi-infinite structure is $200$ -- $2100$~Hz. In order to imitate an infinite structure, all walls have to be totally reflective. At the same time, symmetry of the unit cell and of field distributions within it have to be preserved. Therefore, we select the size of the cross-section such that it's width and height are smaller than the smallest wavelength in the spectrum of a generated signal. Otherwise, the walls have to be perfectly symmetric and elements within the waveguide have to be centered ideally, which is hardly achievable.
In order to increase the reliable frequency range in the waveguide below $\lambda / 2$, we therefore only use half instead of full resonators. Such modification is valid since it does not violate symmetry of the structure. The resonators were manufactured by 3D printing (PLA filament). Note, that to simplify placement of resonators inside the waveguide, there are small place-holders in the slits of the resonators.

All measurements were conducted for generated chirped signal with the frequency changing from $200$ to $2100$~Hz. For each measurement the signal was generated several times and then the measured spectra were averaged in Fourier space. Sampling frequency was always fixed at \unit[44100]{Hz}.
Following the conventional definition, the transmission was defined as
\begin{equation}
    T = 20 \log_{10} \left(p_{\mathrm{tr}}/p_{\mathrm{ref}} \right)
\end{equation}
where $p_{\mathrm{ref}}$ is a pressure amplitude of the wave in the absence of a structure (reference signal) and $p_{\mathrm{tr}}$ is a pressure amplitude of the transmitted wave.

The measurement of transmission for the finite-size chamber were conducted in the same waveguide, as for the case of semi-infinite structure. However, the distance between the walls was increased to $650$~mm. The walls and the ends were covered by foam rubber pyramids (with height $50$~mm and square base $50$ by $50$~mm).
The chamber was located at the center of the waveguide such that microphone (BOYA BY-M1) was placed inside the structure (at the center). All other elements and connections between them were the same as for the semi-infinite system.
The procedure of signal generation and averaging also remained the same, however the frequency range was changed to be from $1500$ to $16500$~Hz. 

\bibliographystyle{naturemag}
\bibliography{bibliography.bib}

\section{Acknowledgements}
The authors thank Aleksandr Kalganov for the help with design and assembly of the experimental setup and Mikhail Kuzmin for sample fabrication. The authors also thank Nikita Olekhno for fruitful discussions and suggestions. This research was supported by Priority 2030 Federal Academic Leadership Program.

\section{Author contributions}
M.K. designed the structure, experimental samples, and the experimental setup, and performed experimental measurements. S.K. and M.K. provided numerical calculations. S.K. participated in the experimental setup assembly and measurements. 
A.M., Y.B., S.M., and D.P. provided guidance on all aspects of the work. A.B. suggested the idea and supervised the project. All authors contributed to writing and editing of the manuscript.

\section{Competing interests}
The authors declare no competing interests.

\newpage
\appendix
\onecolumngrid

\begin{center}
\textbf{\large Supplementary Information}
\end{center}

\section{Coupled Helmholtz resonators}
\label{sec:Coupling}

\begin{figure*}[htbp!]
    \centering
    \includegraphics[width=\linewidth]{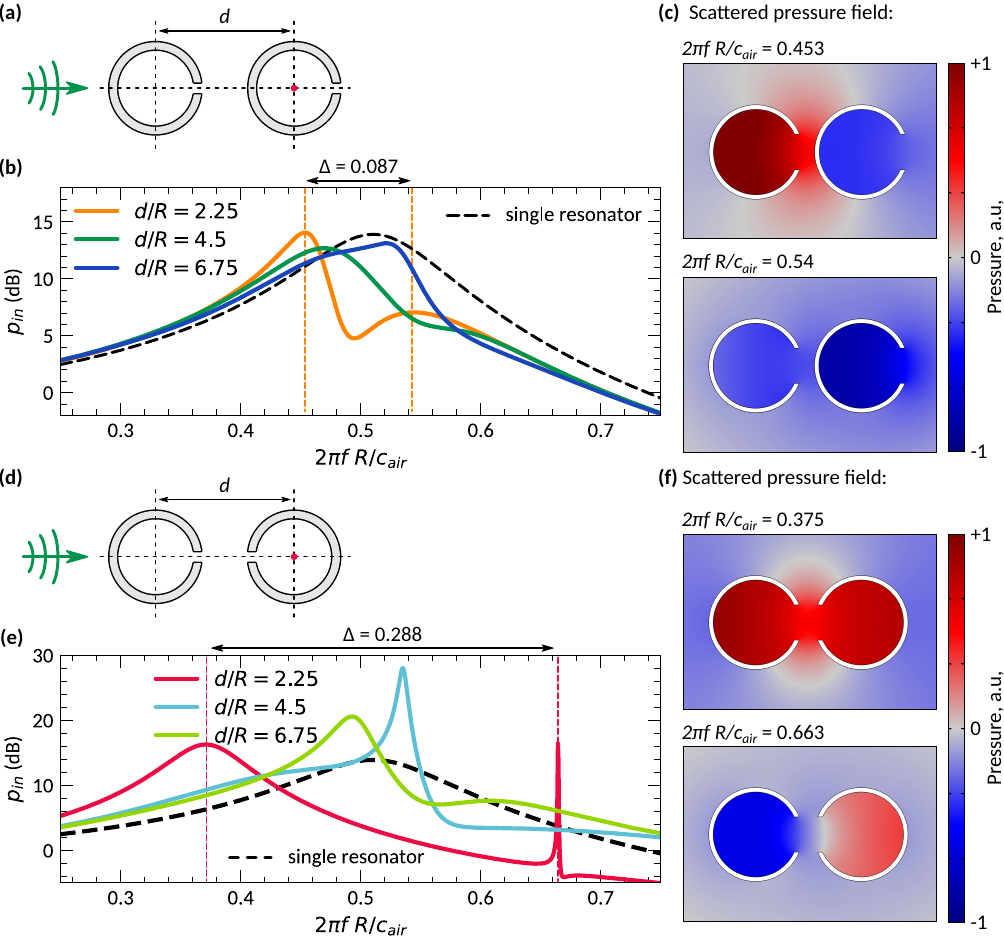}
    \caption{{\bfseries Coupling between two resonators.} The system consists of a pair of the Helmholtz resonators located in a free space, such that the incident pressure field propagates along the $x$-axis. Two cases are considered, namely (a) the CC-like geometry and (d) the strongly coupled resonators. Panels (b) and (e) shows corresponding frequency spectra calculated at the sampling point located at the center of the rightmost resonator (the red dot). The spectral distances between the split resonances for the case of $d/R = 2.25$ are indicated by $\Delta$. Distributions of the scattered pressure field for the case of $d/R = 2.25$ are shown in panels (c) and (f); the frequencies correspond to the resonances indicated by the vertical dashed lines on the spectra.}
    \label{fig:coupling}
\end{figure*}

Since it is claimed that the broadband noise-insulation is achieved due to the strong local coupling of the Helmholtz resonators we would like to provide a proof that the coupling is indeed strong. For that we consider a pair of the resonators in a free space and calculate pressure inside one of them, assuming that the incident wave with the amplitude $p_0 = \unit[1]{Pa}$ propagates along the $x$-axis [see Figs.~\ref{fig:coupling}(a) and~\ref{fig:coupling}(d)]. The resulting spectra are shown in Figs.~\ref{fig:coupling}(b) and ~\ref{fig:coupling}~(e). The splitting of the Helmholtz resonance decreasing with the increase of the distance between the resonators is the manifestation of the coupling between the resonators. For the case of the considered geometry [Fig.~\ref{fig:coupling}(d)], the splitting is much larger than for the CC-like one [Fig.~\ref{fig:coupling}(a)]. In addition, the field distribution in the proposed geometry is fundamentally different from the CC-like geometry since it has additional mirror symmetry with respect to $y$-axis. Hence, the split Helmholtz resonance occurs as in-phase and out-of-phase excitation of the resonators characterized by symmetric and anti-symmetric field distributions, respectively [see Figs.~\ref{fig:coupling}(c) and~\ref{fig:coupling}(f)]. 

\begin{figure*}[htbp!]
    \centering
    \includegraphics[width=\linewidth]{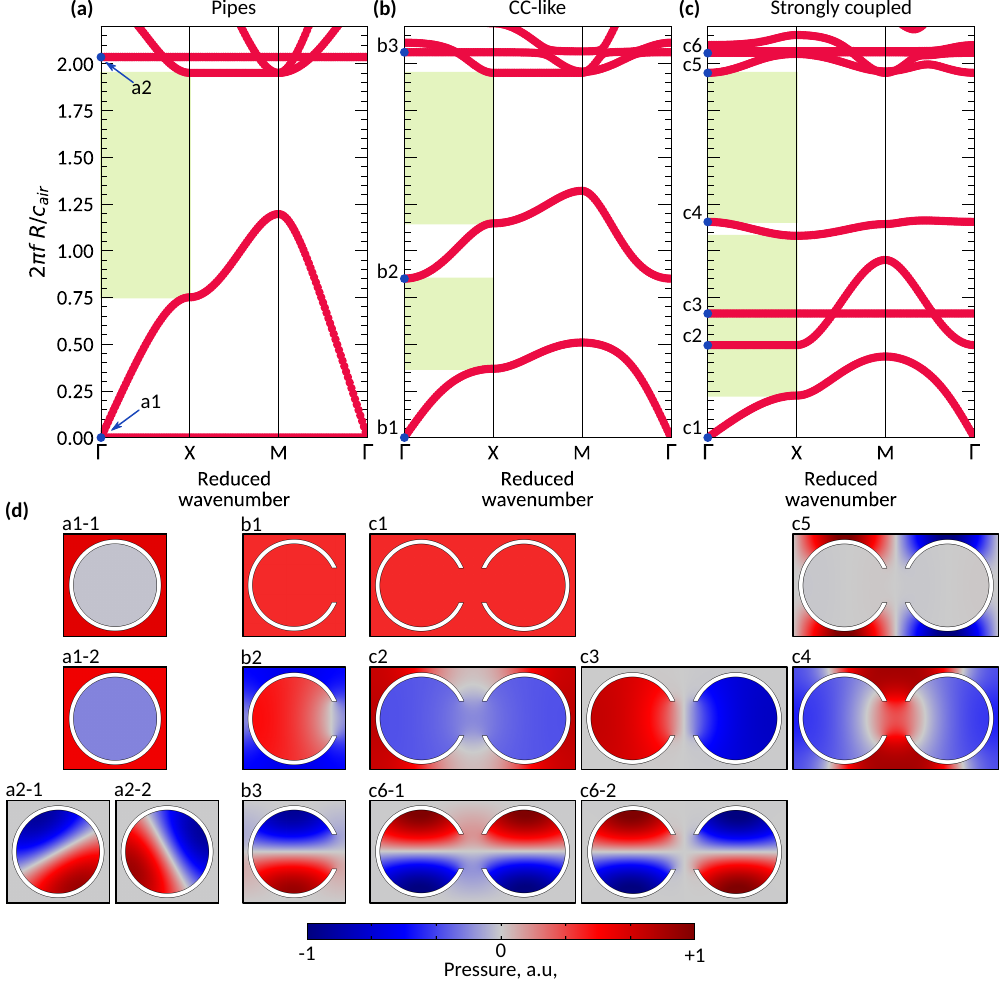}
    \caption{\textbf{Deformation of band structure}. Band diagrams for the case of (a) pipes, (b) CC-like geometry (c) strongly coupled resonators. Shaded green areas indicate the $\Gamma$-$X$ band-gaps. The following parameters of the structures are used: (b) $w/R = 0.75$; (c) $w/R = 0.75$, $d/R = 2.25$; for each panel $r/R = 0.905$. Field distributions shown in panel (d) correspond to the labeled blue points in the band diagrams.}
    \label{fig:eigenmodes}
\end{figure*}

In the case of an infinite arrays, the local coupling results in the pronounced increase of the integral width of band-gaps~\cite{hu2017acoustic}. Figure~\ref{fig:eigenmodes} demonstrates the difference between the band structures for the infinite arrays of resonators with zero slits (which are just circular pipes), CC-like geometry and strongly coupled resonators. Indeed, the introduction of slits to the pipes results in a formation of an additional band [see Fig.~\ref{fig:eigenmodes}(b)] as well as the additional band-gap, as consequence. Further, an enhancement of the local coupling leads to the formation of two more $\Gamma$-$X$ band-gaps. In this case, the out-of-phase excitation of the Helmholtz resonance corresponds to the flat band indicating strong localization of the field inside the resonators as well as the absence of coupling between the neighboring couples. In turn, the other mode, which is flat only within the $\Gamma X$ interval, is associated with the in-phase excitation of the Helmholtz resonance.

To trace the deformation of band structure caused by introduction of slit and the consequent enhancement of the local coupling, we consider field distributions within the unit cells (at $\Gamma$-point). For the structure made of pipes there are two modes with zero frequency. One of them, labelled as a1-1 in Fig.~\ref{fig:eigenmodes}(d), is a structural resonance, typical for periodic structures. This mode is present in both other band diagrams [b1 in Fig.~\ref{fig:eigenmodes}(b) and c1 in Fig.~\ref{fig:eigenmodes}(c)]. Another mode, labelled as a1-2, is the flat band with zero frequency actually corresponding to the eigenmode of a Helmholtz resonator with zero slit. As soon as the slit is introduced, this band deforms and shifts to the area of larger frequencies. Hence, the mode a1-2 of pipe arrays can be associated with the mode b2, which is a Helmholtz resonance of a CC-like structure. When the coupling is enhanced, this mode splits on two [c2 and c3 in Fig.~\ref{fig:eigenmodes}(c)]. In this case, the field distribution patterns within a single unit cell of the periodic structure are similar to those observed for a single pair of resonators presented in Fig.~\ref{fig:coupling}(f). The anti-symmetric mode, labelled as c3  is perfectly localized within the resonators meaning that there is no coupling between neighbouring pairs, which is the reason why the corresponding band is flat. Also, there is another additional mode c4, which may be considered as the additional structural resonance resulting from the deformation of the unit cell. 

Another type of modes are the modes of a circular pipe, which also correspond to flat bands in each of the diagrams. For the case of pipes, these are two degenerate modes a2-1 and a2-2. When the slit is introduced, the degeneracy lifts due to the break of symmetry, but the mode with similar field distribution remains in the system. Then, the structure with the strong coupling is characterized by two similar nearly-degenerate modes (c6-1 and c6-2). Finally, the mode c5 is the structural resonance, such that similar resonances are present in all of the considered structures, but at higher frequencies.

\newpage
\section{Ventilation properties of the structures}
\label{sec:Ventilation}

\begin{figure*}[htbp!]
    \centering
    \includegraphics[width=\linewidth]{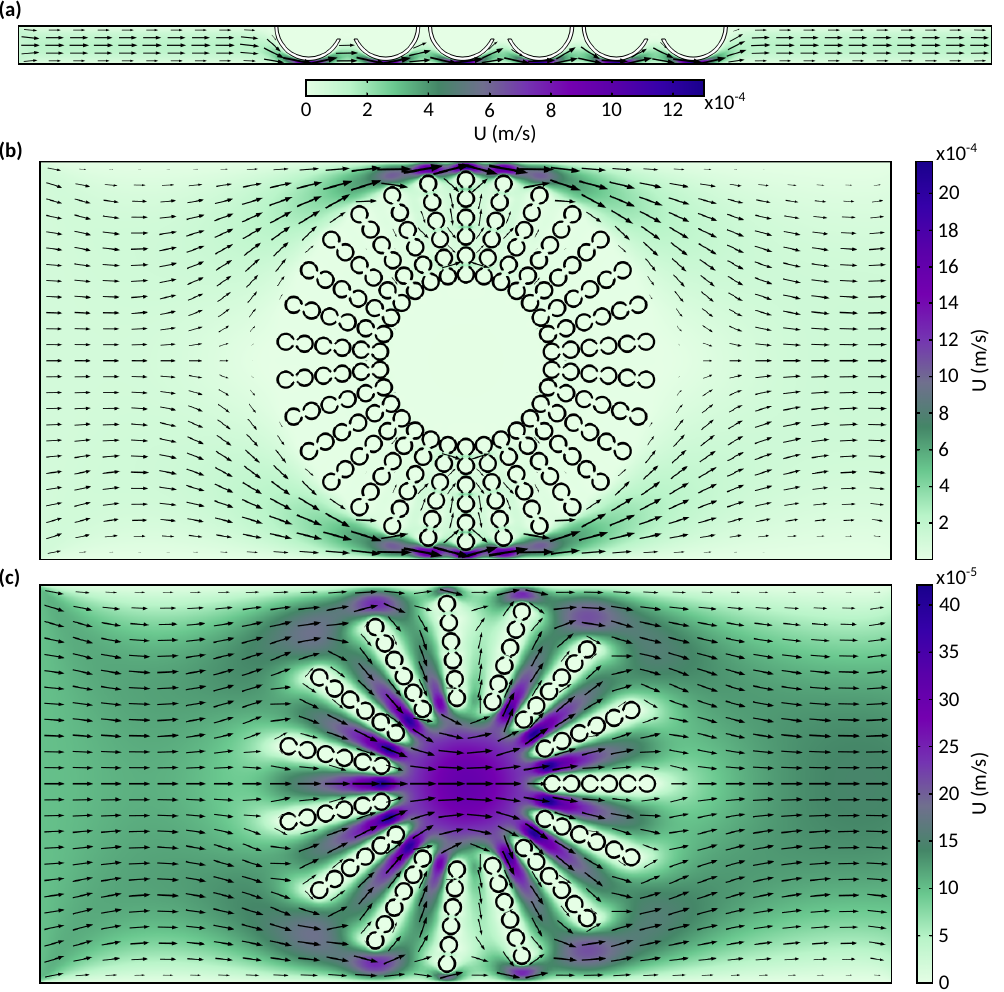}
    \caption{{\bfseries Air flow through the designed structures.} (a) Semi-infinite structure in the duct, (b) \textit{metahouse} chamber and (c) sparse \textit{metahouse} chamber considered in the main text. For each panel left boundary is considered to be an inlet port with normal inflow velocity $U_0 = \unit[10^{-4}]{m/s}$. The direction of the flow is shown by black arrows while the color indicates its velocity $U$.}
    \label{fig:air_flow}
\end{figure*}

In order to demonstrate the ventilation properties of the proposed structures, we calculate the laminar air flow with the help of COMSOL Multiphysics software package. For all cases, the normal inflow velocity was taken to be $U_{0} = \unit[10^{-4}]{m/s}$ and the no-slip wall boundary condition was used. The results shown in Fig.~\ref{fig:air_flow}(a) demonstrate that the semi-infinite structure does not violate the air flow, suggesting that the structure might be suitable for noise insulation of ventilated ducts.

\begin{figure*}[htbp!]
    \centering
    \includegraphics[width=\linewidth]{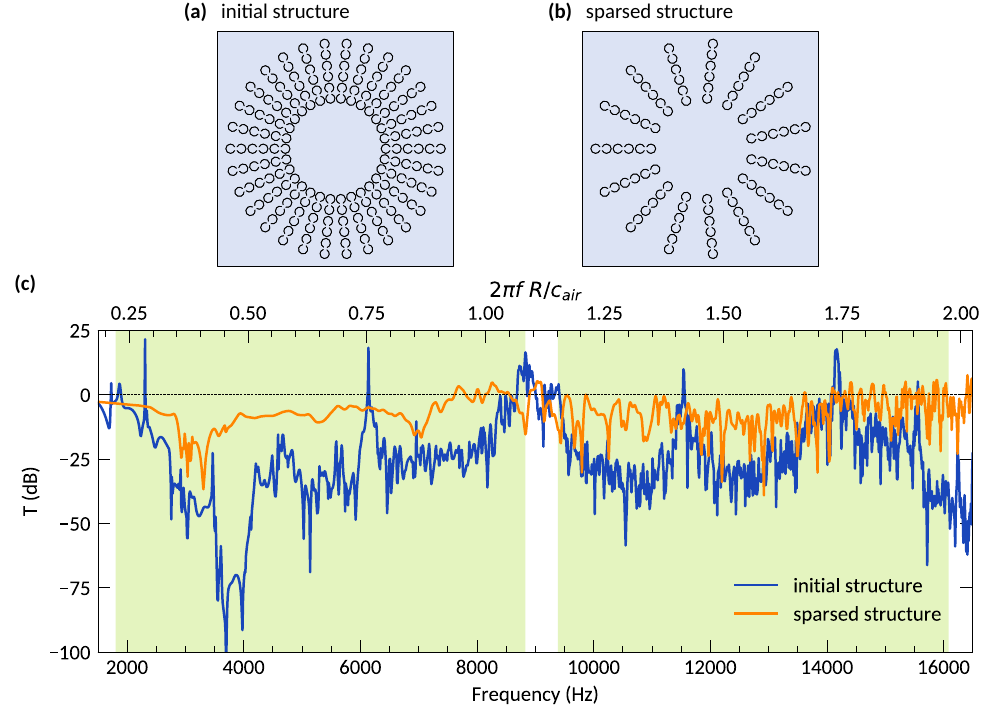}
    \caption{{\bfseries Transmission of the sparse \textit{metahouse} chamber.} Schematic representation of the (a) initial [see Fig.~6 from the main text] and (b) sparse structures. (c) Simulated transmission spectra for both cases. Average transmission for the sparse structure is $-7$~dB and for the initial structure is $-25$~dB.}
    \label{fig:metacage_sparsed}
\end{figure*}

Switching to the case of \textit{metahouse} chamber,  Fig.~\ref{fig:air_flow}(b) shows that the laminar flow inside the structure is quite weak. Flow velocity at the center of the chamber is about $\unit[10^{-5}]{m/s}$, which is one order smaller than the inflow velocity. Hence, the initial structure for which the experimental measurements were done is poorly ventilated. At the same time, considering the use of such structures in the real-world conditions of urban parks, it can be hypothesised that air will definitely penetrate inside the chamber which at the same time may provide some protection from wind.
Increasing the sparseness enhances the air flow inside the structure, as is seen in Fig.~\ref{fig:air_flow}(c). However, in this case the flow velocity inside the chamber is higher than the inlet one. The reason for this is the cone-like narrowing of the channels in-between the arrays of the resonators resulting in the flow acceleration. 
While the sparsed structure has a half of the initial number of the resonators, its noise-insulating properties are still quite reasonable. In this case the numerically calculated average transmission is about $-7$~dB as the Fig.~\ref{fig:metacage_sparsed} shows.
The conclusion which can be made from the Figs.~\ref{fig:air_flow} and~\ref{fig:metacage_sparsed} is that it is possible to find a compromise between the ventilating and noise-insulating properties of the proposed \textit{metahouse} structure by varying its sparseness.

\end{document}